\documentclass[aps,prd,twocolumn, groupedaddress]{revtex4-2}

\usepackage{amsmath}
\usepackage{amssymb}
\usepackage{amsfonts}
\usepackage{bm, bbm}
\usepackage{tensor}

\usepackage{url}

\usepackage{breakurl}
\usepackage[breaklinks]{hyperref}

\newcommand{\bfuncs}{$\beta$-functions}

\newcommand{\msbar}{$\overline{\mathrm{MS}}$}

\newcommand*{\trans}{\mathrm{T}}

\newcommand{\phid}{\phi_{\hat{d}}}

\newcommand{\hs}{\hspace{.033cm}}

\newcommand{\tr}[1]{\mathrm{Tr}\left[ #1 \right]}

\newcommand{\fc}[2]{\bm{\mathfrak{\xi}}_{#2}^{\,(#1)}\,}
\newcommand{\tc}[2]{\bm{\mathfrak{\tau}}_{#2}^{(#1)}\,}
\newcommand{\sm}[2]{\bm{\mathfrak{\sigma}}_{#2}^{(#1)}\,}

\newcommand{\y}[1]{y_{#1}}
\newcommand{\ya}{\y{a}}
\newcommand{\yb}{\y{b}}
\newcommand{\yc}{\y{c}}
\newcommand{\yd}{\y{d}}
\newcommand{\ye}{\y{e}}

	\newcommand{\yt}[1]{\tilde{y}_{#1}}
	\newcommand{\yta}{\yt{a}}
	\newcommand{\ytb}{\yt{b}}
	\newcommand{\ytc}{\yt{c}}
	
	\newcommand{\yte}{\yt{e}}
	\newcommand{\ytf}{\yt{f}}
\newcommand{\m}{m\hs}
	\newcommand{\mt}{\tilde{m}\hs}

\newcommand{\G}[1]{G^2_{#1 \,}}

\newcommand{\CF}{C_2(F)\hs}
	\newcommand{\CFt}{\tilde{C}_2(F)\hs}
\newcommand{\YF}{Y_2(F)\hs}
	\newcommand{\YFt}{\tilde{Y}_2(F)\hs}
\newcommand{\CFG}{C_2(F,G)\hs}
\newcommand{\CFS}{C_2(F,S)\hs}
\newcommand{\CFF}{C_2(F,F)\hs}
\newcommand{\YFCF}{Y_2(F,C_F)\hs}
\newcommand{\YFCS}{Y_2(F,C_S)\hs}

	\newcommand{\YFYFt}{\tilde{Y}_2(F,Y_F)\hs}
\newcommand{\YFYS}{Y_2(F,Y_S)\hs}
\newcommand{\YfF}{Y_4(F)\hs}

\newcommand{\CS}[2]{\left[C_2(S)\right]_{#1 #2}}
\newcommand{\YS}[2]{\left[Y_2(S)\right]_{#1 #2}}
\newcommand{\CSG}[2]{\left[C_2(S, G)\right]_{#1 #2}}
\newcommand{\CSS}[2]{\left[C_2(S, S)\right]_{#1 #2}}
\newcommand{\CSF}[2]{\left[C_2(S, F)\right]_{#1 #2}}
\newcommand{\YSCF}[2]{\left[Y_2(S, C_F)\right]_{#1 #2}}
\newcommand{\YSYF}[2]{\left[Y_2(S, Y_F)\right]_{#1 #2}}
\newcommand{\YfS}[2]{\left[Y_4(S)\right]_{#1 #2}}

\begin{document}


\title{General RGEs for dimensionful couplings in the $\overline{\mathrm{MS}}$ scheme}


\author{Lohan Sartore}
\email[]{sartore@lpsc.in2p3.fr}
\affiliation{Laboratoire de Physique Subatomique et de Cosmologie, Universit\'e Grenoble-Alpes, CNRS/IN2P3, 53 Avenue des Martyrs, 38026 Grenoble, France}

\begin{abstract}
Recently, Poole \& Thomsen have presented the Renormalization Group Equations (RGEs) in the \msbar~scheme for dimensionless parameters of a general renormalizable gauge theory in a new formalism based on the Local Renormalization Group. In this Letter, we apply the dummy field method to the expressions in this formalism in order to derive the RGEs for the dimensionful couplings which so far have been missing in this approach. The complete set of RGEs has been implemented in a new version of the public code PyR@TE dedicated to the automatic computation of the RGEs for any general renormalizable (non-supersymmetric) gauge theory.
\end{abstract}


\maketitle

\emph{Introduction.}~--- The knowledge of Renormalization Group Equations (RGEs) is crucial in the context of studies of gauge theories since they provide the necessary link between the couplings of the theory at different energy scales. The expressions for the dimensionless coulings RGEs of a general gauge theory were first presented almost forty years ago \cite{machacek_two-loop_I_1983, machacek_two-loop_II_1984, machacek_two-loop_III_1985, jack_two-loop_1982, Jack_1983, jack_general_1985}. These results were later extended to the case of dimensionful parameters \cite{luo_two-loop_2003} using the so-called \textit{dummy field method} \cite{martin_two-loop_1994, luo_two-loop_2003}, which relates the \bfuncs~of dimensionful parameters to those of dimensionless ones. A thorough presentation of the dummy field method is given in \cite{schienbein_revisiting_2019} along with a proper derivation of the general two-loop RGEs for dimensionful couplings in the \msbar~scheme, correcting some of the results presented in \cite{luo_two-loop_2003}. Using these general expressions by hand can be quite involved and error-prone. For this reason, they have been implemented in the Mathematica package SARAH \cite{staub_sarah_2014} and the Python package PyR@TE \cite{lyonnet_pyrte_2014, lyonnet_pyrte_2017}, which are both publicly available.

Recently, a new formalism was proposed \cite{poole_constraints_2019} for the RGEs of general gauge theories that possesses many interesting features compared to the old Machacek \& Vaughn formalism, both on the conceptual and practical sides. Within the framework of the Local Renormalization Group \cite{OSBORN198997, JACK1990647, OSBORN1991486}, relations can be derived that involve the coefficients appearing in the \bfuncs~of gauge, Yukawa and quartic couplings at different loop orders: these are the so-called Weyl Consistency Conditions. Based on these relations and the existing results in the literature at two-loop, the authors of \cite{poole_constraints_2019} were able to present for the first time the general expressions of the gauge coupling \bfuncs~at the three-loop order. More recently, this formalism helped in the derivation of the 4-loop gauge \bfuncs~in the Standard Model \cite{davies_gauge_2020}. In the light of these successful developments, we believe that this formalism will allow in a near future further progress in the field of general RGEs.

Compared to the previous formalism, several new attractive features should be noted. First, the expressions of the \bfuncs~are equally concise for gauge theories based either on a simple gauge group or a semi-simple one. In particular, in the presence of multiple Abelian gauge factors, the effects of kinetic mixing \cite{luo_renormalization_U1_2003, fonseca_renormalization_2013} are encompassed in the formalism in a systematic way. Another salient feature is the easy adaptability of the formalism to renormalization schemes other than \msbar, the Weyl Consistency Conditions being scheme-independent relations. Finally, it was shown \cite{poole_constraints_2019, poole_weyl_2019} that the Weyl consistency conditions can help solve the ambiguities related to the treatment of $\gamma_5$ in fermion loops starting to appear at the 3- and 4-loop order in Yukawa and gauge \bfuncs, respectively.

Currently, there is however one missing piece compared to the previous formalism, namely the expressions of the RGEs for all (renormalizable) dimensionful coupings. The aim of the present Letter is to bridge this gap providing general expressions for the \bfuncs~of dimensionful couplings, obtained by properly applying the dummy field method to the formalism of Poole \& Thomsen.\\

\emph{Description of the formalism.}~--- We consider a general gauge theory containing an arbitrary set of real scalars $\phi_a$ and Weyl fermions $\psi_i$. Following \cite{poole_constraints_2019}, we define the Majorana spinor
\begin{equation}\label{majoranaSpinor}
\tensor{\Psi}{_i} \equiv \begin{pmatrix}
	\psi \\
	\psi^\dagger
\end{pmatrix}_i \,.
\end{equation}
With this definition, the most general Lagrangian density may be written (see \cite{poole_constraints_2019} for details):
\begin{align}
\begin{split}\label{fullLag}
	\mathcal{L} = &-\frac{1}{4}G^{\,-2}_{AB} F^A_{\mu\nu} F^{B\,\mu\nu} + \frac{1}{2} (D_\mu \phi)_a (D^{\,\mu} \phi)_a \\
	& + \frac{i}{2} \Psi^\trans \begin{pmatrix} 0 & \sigma^{\mu} \\ \bar{\sigma}^{\mu} & 0\end{pmatrix} D_\mu\Psi\\
	&-\frac{1}{2} \tensor{y}{_a_i_j} \Psi_i \Psi_j\,\phi_a - \frac{1}{2} \tensor{m}{_i_j} \Psi_i \Psi_j \\
	& -\frac{1}{2} \tensor{\mu}{_a_b} \phi_a \phi_b -\frac{1}{3!} \tensor{t}{_a_b_c} \phi_a \phi_b \phi_c -\frac{1}{4!} \tensor{\lambda}{_a_b_c_d} \phi_a \phi_b \phi_c \phi_d \, ,
\end{split}
\end{align}
where $G_{AB}$ is a matrix containing the gauge couplings associated to the (semi-)simple gauge group of the theory, which can be decomposed as:
\begin{equation}
	\mathcal{G} = \prod_{u=1}^{N}\, \mathcal{G}_u\,.
\end{equation}
Letting $d_u$ be the dimension of the gauge factor $\mathcal{G}_u$, the covariant derivatives for fermions and scalars are respectively defined as:
\begin{align}
	D_\mu \Psi_i &= \partial_\mu \Psi_i - i \sum_{u=1}^{N} \sum_{A_u = 1}^{d_u} V^{A_u}_\mu \left(T^{A_u}\right)_{ij} \Psi_j\,,\label{CDf}\\
	D_\mu \phi_a &= \partial_\mu \phi_a - i \sum_{u=1}^{N} \sum_{A_u = 1}^{d_u} V^{A_u}_\mu \left(T^{A_u}_\phi\right)_{ab} \phi_b\,\label{CDs}.
\end{align}
The notation may be simplified using the summation convention
\begin{equation}
	\sum_A = \sum_u\,\sum_{A_u}^{d_u}
\end{equation}
for the gauge indices. Defining
\begin{equation}
	n = \sum_{u=1}^{N} d_u\,,
\end{equation}
the covariant derivatives~\eqref{CDf} and~\eqref{CDs} may be rewritten as
\begin{align}
	D_\mu \Psi_i &= \partial_\mu \Psi_i - i \sum_{A = 1}^{n} V^{A}_\mu \left(T^{A}\right)_{ij} \Psi_j\,,\\
	D_\mu \phi_a &= \partial_\mu \phi_a - i \sum_{A = 1}^{n} V^{A}_\mu \left(T^{A}_\phi\right)_{ab} \phi_b\,.
\end{align}
In a theory with $p$ Abelian gauge factors, the gauge coupling matrix $G^2$ takes the general form
\begin{equation}
	G^2 = \begin{pmatrix}
				h_{11} & \cdots & h_{1p} & & \\
				\vdots & \ddots & \vdots &  & \\
				h_{p1} & \cdots & h_{pp} &  & \\
				       &        &        &  g_{p+1}^2 & \\
				       &        &        &            & \ddots \\
				       &        &        &            &          & g_n^2\;
			\end{pmatrix}
			\equiv 
			\begin{pmatrix}
				H^2 & 0 \\
				0 & G_{\mathrm{na}\,}^2
			\end{pmatrix}\,.
\end{equation}
The diagonal entries of $G_{\mathrm{na}\,}^2$ are populated by the non-Abelian gauge couplings according to
\begin{equation}
	\left(G_{\mathrm{na}\,}^2\right)_{AB} = g_u^2 \, \delta^{A_uB_u}\,,\quad u > p
\end{equation}
and the matrix $H^2$ can be decomposed as
\begin{equation}
	H^2 = G_{\mathrm{mix}} G_{\mathrm{mix}}^\trans\,,
\end{equation}
where $G_{\mathrm{mix}}$ is a $p\times p$ matrix developing non-zero off-diagonal components in presence of kinetic mixing.\\

The \bfuncs~of the dimensionless couplings of the model -- namely gauge, Yukawa and quartic couplings -- are defined as follows \cite{poole_constraints_2019}:
\begin{align}
	\beta_{AB} \equiv \frac{d G_{AB}^2}{d t} &= \frac{1}{2} \sum_{\mathrm{perm}} \sum_\ell \frac{1}{\left(4\pi\right)^{2\ell}}\, G^2_{AC\,} \beta^{(\ell)}_{CD\,} G^2_{DB}\,,\label{gaugeBeta}\\
		\beta_{a i j} \equiv \frac{d y_{a i j}}{d t} &= \frac{1}{2} \sum_{\mathrm{perm}} \sum_\ell \frac{1}{\left(4\pi\right)^{2\ell}}\, \beta^{(\ell)}_{a i j}\,,\\
		\beta_{a b c d} \equiv \frac{d \lambda_{a b c d}}{d t} &= \frac{1}{4!} \sum_{\mathrm{perm}} \sum_\ell \frac{1}{\left(4\pi\right)^{2\ell}}\,\beta^{(\ell)}_{a b c d}\,\label{quarticBeta},
\end{align}
with $\ell$ denoting the perturbative loop-order. We generalize this definition to the dimensionful couplings of the model (fermion mass, scalar trilinear and scalar mass couplings respectively):
{\allowdisplaybreaks
\begin{align}
		\beta_{i j} \equiv \frac{d m_{i j}}{d t} &= \frac{1}{2} \sum_{\mathrm{perm}} \sum_\ell \frac{1}{\left(4\pi\right)^{2\ell}}\, \beta^{(\ell)}_{i j}\,,\\
		\beta_{a b c} \equiv \frac{d t_{a b c}}{d t} &= \frac{1}{3!} \sum_{\mathrm{perm}} \sum_\ell \frac{1}{\left(4\pi\right)^{2\ell}}\,\beta^{(\ell)}_{a b c}\,,\\
		\beta_{a b} \equiv \frac{d \mu_{a b}}{d t} &= \frac{1}{2} \sum_{\mathrm{perm}} \sum_\ell \frac{1}{\left(4\pi\right)^{2\ell}}\,\beta^{(\ell)}_{a b}\,.
\end{align}
}

In \cite{poole_constraints_2019}, the expressions of the dimensionless $\beta^{(\ell)}$ are given as a sum of individual contributions, each one associated with a particular diagram and weighted by a renormalization scheme-dependent coefficient.\\

\emph{The dummy field method.}~--- We consider a general theory containing only dimensionless couplings, for which the expressions of the RGEs are known. In our notation, the non-kinetic part of the associated Lagrangian density is given by
\begin{equation}\label{dummyLag1}
		\mathcal{L}_0 = -\frac{1}{2} \tensor{y}{_a_i_j} \Psi_i \Psi_j\,\phi_a  -\frac{1}{4!} \tensor{\lambda}{_a_b_c_d} \phi_a \phi_b \phi_c \phi_d \, .
\end{equation}
The theory is then extended with a non-propagating scalar field with no gauge interactions. Reusing the notation from \cite{schienbein_revisiting_2019}, this \emph{dummy} field is denoted $\phid$ and satisfies $D_\mu \phid = 0$. Making explicit the terms involving $\phid$ and discarding a constant as well as a term linear in $\phid$, the Lagrangian~\eqref{dummyLag1} may be rewritten
\begin{align}
\begin{split}\label{dummyLag2}
	\mathcal{L}_0 = &-\frac{1}{2} \tensor{y}{_a_i_j} \Psi_i \Psi_j\,\phi_a -\frac{1}{4!} \tensor{\lambda}{_a_b_c_d} \phi_a \phi_b \phi_c \phi_d \\
	& - \frac{1}{2} \tensor{y}{_{\hat{d}}_i_j} \Psi_i \Psi_j\,\phid -\frac{1}{3!} \tensor{\lambda}{_a_b_c_{\hat{d}}} \phi_a \phi_b \phi_c \phid \\
	&  -\frac{1}{4} \tensor{\lambda}{_a_b_{\hat{d}}_{\hat{d}}} \phi_a \phi_b \phid \phid;\,.
\end{split}
\end{align}

The above form makes it clear that the Lagrangian~\eqref{fullLag} of a theory containing dimensionful coupling can be recovered if the following identifications are made:
\begin{equation}\label{dummyMapping}
	\tensor{y}{_{\hat{d}}_i_j} \phid = \tensor{\m}{_i_j} \,,\quad
	\tensor{\lambda}{_a_b_{\hat{d}}_{\hat{d}}} \phid \phid = 2 \tensor{\mu}{_a_b}\,,\quad
	\tensor{\lambda}{_a_b_c_{\hat{d}}} \phid = \tensor{t}{_a_b_c}\,.
\end{equation}

The mapping~\eqref{dummyMapping} allows for a straightforward derivation of the individual contributions to the \bfuncs~of the dimensionful couplings, starting from the known expressions in the dimensionless case. For instance, the diagrams contributing to the fermion mass RGEs ($\beta^{(\ell)}_{i j}$) are obtained from the individual contributions to the Yukawa couplings RGEs ($\beta^{(\ell)}_{a i j}$), when the external scalar field $\phi_a$ is taken to be the dummy field $\phid$. Consequently, one should in practice perform the following replacements:
\begin{equation*}
	a\rightarrow\hat{d},\ \ \tensor{y}{_a_i_j} \rightarrow \tensor{\m}{_i_j},\ \  \tensor{\lambda}{_a_b_c_d} \rightarrow \tensor{t}{_a_b_c}\,.
\end{equation*}

Similarly, starting from the quartic couplings RGEs, one may derive the the \bfuncs~for the trilinear and scalar mass couplings when one	or two of the external scalar legs are replaced by a dummy field. This procedure is applied on a diagrammatic basis, allowing in particular the identification of unphysical tadpole contributions which must be discarded \cite{schienbein_revisiting_2019}. \\

\emph{Results.}~--- We now turn to the presentation of the results. The various quantities appearing in the expressions of the \bfuncs~are directly taken from \cite{poole_constraints_2019}. Therefore, we invite the reader to refer to the definitions presented therein. 
We show below the whole set of RGEs obtained in the \msbar~scheme for the fermion mass, trilinear and scalar mass couplings. The following equations complete the list of RGEs presented in \cite{poole_constraints_2019} for the dimensionless couplings. We use a notation where the fermion indices are made implicit. In this context, $m$, $y_a$, and any other tensor carrying two fermion indices may be seen as matrices in the space of the fermions of the theory.

\begin{widetext}

\begin{center} \textbf{Fermion mass \bfuncs} \end{center}

{\allowdisplaybreaks
At 1-loop: 
\begin{equation}
	\beta^{(1)} = \fc{1}{1} \m C_2(F)\enspace + \enspace \fc{1}{2} \yb\m \yb \enspace + \enspace \fc{1}{3} \m \YFt\,.
\end{equation}

At 2-loop:
\begin{align}
	\beta^{(2)} ={}& \fc{2}{1} \CFt \m \CF &&+ \fc{2}{2} \m\CF\CF &&+ \fc{2}{3} \m \CFG \nonumber\\
	+\ & \fc{2}{4} \m \CFS  &&+  \fc{2}{5} \m\CFF   &&+  \fc{2}{6} \yb T^A \ytb \m T^B \G{AB} \nonumber\\
	+\ & \fc{2}{7} \YF \tilde{T}^A \m T^B \G{AB}   &&+  \fc{2}{8} \yb \mt \yc \CS{b}{c}  &&+  \fc{2}{9} \yb\CF\mt\yb \nonumber\\
	+\ & \fc{2}{10} \yb \mt \yb \CF   &&+  \fc{2}{11} \YFCS \m  &&+  \fc{2}{12} \YFCF \m \nonumber\\
	+\ & \fc{2}{13} \m \YFt \CF   &&+  \fc{2}{14} \yb\ytc\yd\, t_{bcd}  &&+  \fc{2}{15} \yb\ytc\m\ytb\yc \nonumber\\
	+\ & \fc{2}{16} \yb \mt \yc \ytb \yc  &&+  \fc{2}{17} \yb \ytc \m \ytc \yb  &&+  \fc{2}{18} \YfF \m \nonumber\\
	+\ & \fc{2}{19} \yb \mt \YF \yb  &&+  \fc{2}{20} \m \YFYFt  &&+  \fc{2}{21} \yb \mt \yc \YS{b}{c} \nonumber\\
	+\ & \fc{2}{22} \YFYS \m\,.
\end{align}
}

\begin{center} \textbf{Trilinear couplings \bfuncs} \end{center}

{\allowdisplaybreaks
At 1-loop: 
\begin{equation}
	\beta^{(1)}_{abc} = \tc{1}{1} \CS{a}{e} t_{ebc} \  + \  \tc{1}{2} \lambda_{abef} t_{efc} \  + \  \tc{1}{3} \YS{a}{e} t_{ebc} \  + \  \tc{1}{4} \tr{\m\yta\yb\ytc}\,.
\end{equation}

At 2-loop:
\begin{align}
	\beta^{(2)}_{abc} ={}& \tc{2}{1} (T_\phi^A T_\phi^C)_{ae} \G{AB}\G{CD}( T_\phi^B T_\phi^D)_{bf\,} t_{efc}  &&+ \tc{2}{2} (T_\phi^A T_\phi^C)_{ab} \G{AB}\G{CD}( T_\phi^B T_\phi^D)_{ef\,} t_{efc} \nonumber \\
+& \tc{2}{3} \CS{a}{e} \CS{b}{f} t_{efc} &&+ \tc{2}{4} \CS{a}{e}\CS{e}{f} t_{fbc} \enspace \nonumber\\
	+& \tc{2}{5} \CSG{a}{e} t_{ebc} &&+ \tc{2}{6} \CSS{a}{e} t_{ebc} \nonumber\\
	+& \tc{2}{7} \CSF{a}{e} t_{ebc} &&+ \tc{2}{8} (T_\phi^A)_{ae}(T_\phi^B)_{bf} \G{AB} \lambda_{efgh} t_{ghc} \nonumber\\
	+& \tc{2}{9} \lambda_{abef} \CS{f}{g} t_{egc} &&+ \tc{2}{10} \CS{a}{e} t_{efg} \lambda_{fgbc}  \nonumber\\
	+& \tc{2}{11} \CS{a}{e}\lambda_{ebfg} t_{fgc} &&+ \tc{2}{12} \lambda_{aefg} \lambda_{efgh} t_{hbc} \nonumber\\
	+& \tc{2}{13} t_{aef} \lambda_{eghb} \lambda_{fghc} &&+ \tc{2}{14} \lambda_{abef} \lambda_{eghc} t_{fgh} \nonumber\\
	+& \tc{2}{15} \lambda_{abef} \lambda_{efgh} t_{ghc} &&+ \tc{2}{16} (T_\phi^A T_\phi^C)_{ab} \G{AB} \G{CD} \tr{T^D T^B \mt \yc} \nonumber\\
	+& \tc{2}{17} (T_\phi^A T_\phi^C)_{ab} \G{AB} \G{CD} \tr{T^D T^B \ytc \m} &&+ \tc{2}{18} \YSCF{a}{e} t_{ebc}  \nonumber\\
	+& \tc{2}{19} \CS{a}{e} \YS{e}{f} t_{fbc} &&+ \tc{2}{20} \lambda_{abef} \YS{f}{g} t_{egc}  \nonumber\\
	+& \tc{2}{21} \G{AB} \tr{\m T^A \yta \yb T^B \ytc} &&+ \tc{2}{22} \G{AB} \tr{\ya T^A \mt \yb T^B \ytc} \nonumber\\
	+& \tc{2}{23} \CS{a}{e} \tr{\ye \mt \yb \ytc} &&+ \tc{2}{24} \CS{a}{e} \tr{\ye \ytb \m \ytc}  \nonumber\\
	+& \tc{2}{25} \tr{\m \yta \yb \ytc \CFt} &&+ \tc{2}{26} \tr{\ya \m \yb \ytc \CFt}  \nonumber\\
	+& \tc{2}{27} \tr{\m \yte \ya \ytf} \lambda_{efbc} &&+ \tc{2}{28} \tr{\ya \yte \yb \ytf} t_{efc}  \nonumber\\
	+& \tc{2}{29} \tr{\m \yta\ye\ytf} \lambda_{efbc} &&+ \tc{2}{30} \tr{\ya\ytb\ye\ytf} t_{efc} \nonumber\\
	+& \tc{2}{31} \YfS{a}{e} t_{ebc} &&+ \tc{2}{32} \YSYF{a}{e} t_{ebc}  \nonumber\\
	+& \tc{2}{33} \tr{\m \yta\yb\yte\yc\yte} &&+ \tc{2}{34} \tr{\ya\mt\yb\yte\yc\yte} \nonumber\\
	+& \tc{2}{35} \tr{\ya\ytb\m\yte\yc\yte} &&+ \tc{2}{36} \tr{\m\yta\ye\ytb\yc\yte}  \nonumber\\
	+& \tc{2}{37} \tr{\m\yta\yb\ytc\YF} &&+ \tc{2}{38} \tr{\ya\mt\yb\ytc\YF} \,.
\end{align}
}

\begin{center} \textbf{Scalar mass \bfuncs} \end{center}

{\allowdisplaybreaks
At 1-loop: 
\begin{align}
	\beta^{(1)}_{ab} ={}& \sm{1}{1} \CS{a}{e} \mu_{eb} &&+ \sm{1}{2} \lambda_{abef} \mu_{ef} &&+ \sm{1}{3} t_{aef} t_{efb} \nonumber \\
	+& \sm{1}{4} \YS{a}{e} \mu_{eb} &&+ \sm{1}{5} \tr{\m\mt\ya\ytb} &&+ \sm{1}{6} \tr{\m \yta \m \ytb}\,.
\end{align}

At 2-loop:
\begin{align}
	\beta^{(2)}_{ab} ={}& \sm{2}{1} (T_\phi^A T_\phi^C)_{ae} \G{AB}\G{CD}( T_\phi^B T_\phi^D)_{bf\,} \mu_{ef}  &&+ \sm{2}{2} (T_\phi^A T_\phi^C)_{ab} \G{AB}\G{CD}( T_\phi^B T_\phi^D)_{ef\,} \mu_{ef} \nonumber \\
+& \sm{2}{3} \CS{a}{e} \CS{b}{f} \mu_{ef} &&+ \sm{2}{4} \CS{a}{e}\CS{e}{f} \mu_{fb} \enspace \nonumber\\
	+& \sm{2}{5} \CSG{a}{e} \mu_{eb} &&+ \sm{2}{6} \CSS{a}{e} \mu_{eb} \nonumber\\
	+& \sm{2}{7} \CSF{a}{e} \mu_{eb} &&+ \sm{2}{8} (T_\phi^A)_{ae}(T_\phi^B)_{bf} \G{AB} \lambda_{efgh} \mu_{gh} \nonumber\\
	+& \sm{2}{9} \lambda_{abef} \CS{f}{g} \mu_{eg} &&+ \sm{2}{10} t_{aef} \CS{f}{g} t_{egb} \nonumber \\
	+& \sm{2}{11} \CS{a}{e} \lambda_{ebfg} \mu_{fg}  &&+  \sm{2}{12} \CS{a}{e} t_{efg} t_{fgb} \nonumber\\
	+& \sm{2}{13} \lambda_{aefg} \lambda_{efgh} \mu_{hb} &&+ \sm{2}{14} \lambda_{aegh} \lambda_{bfgh} \mu_{ef} \nonumber\\
	+& \sm{2}{15} \lambda_{abef} t_{egh} t_{fgh} &&+ \sm{2}{16} t_{aef} t_{egh} \lambda_{bfgh} \nonumber \\
	+& \sm{2}{17} \lambda_{abef} \lambda_{efgh} \mu_{gh} &&+ \sm{2}{18} t_{aef}\lambda_{efgh} t_{ghb}  \nonumber \\
	+& \sm{2}{19} (T_\phi^A T_\phi^C)_{ab} \G{AB} \G{CD} \tr{T^D T^B \mt \m} &&+ \sm{2}{20} \YSCF{a}{e} \mu_{eb} \nonumber\\
	+& \sm{2}{21} \CS{a}{e} \YS{e}{f} \mu_{fb} &&+ \sm{2}{22} \lambda_{abef} \YS{f}{g} \mu_{eg}  \nonumber\\
	+& \sm{2}{23} t_{aef} \YS{f}{g} t_{egb} &&+ \sm{2}{24} \G{AB} \tr{\ya T^A \mt \m T^B \ytb}   \nonumber\\
	+& \sm{2}{25} \G{AB} \tr{\ya T^A \ytb \m T^B \mt} &&+ \sm{2}{26} \G{AB} \tr{\ya T^A \mt \yb T^B \mt}   \nonumber\\
	+& \sm{2}{27} \CS{a}{e}\tr{\ye \mt \yb \m} &&+ \sm{2}{28} \CS{a}{e}\tr{\ye \mt \m \yb}  \nonumber\\
	+& \sm{2}{29}  \tr{\ya\ytb\m\mt\CFt} &&+ \sm{2}{30} \tr{\ya\mt\yb\mt\CFt}  \nonumber\\
	+& \sm{2}{31}  \tr{\ya\mt\m\ytb\CFt} &&+ \sm{2}{32} \tr{\m\yta\yb\mt\CFt}  \nonumber\\
	+& \sm{2}{33} \tr{\ya\yte\yb\ytf} \mu_{ef}  &&+ \sm{2}{34} \tr{\ya\yte\m\ytf} t_{efb} \nonumber\\
	+& \sm{2}{35}  \tr{\m\yte\m\ytf} \lambda_{efab}&&+ \sm{2}{36} \tr{\ya\ytb\ye\ytf} \mu_{ef}  \nonumber\\
	+& \sm{2}{37} \tr{\ya\mt\ye\ytf} t_{efb} &&+ \sm{2}{38} \tr{\m\mt\ye\ytf} \lambda_{efab}  \nonumber\\
	+& \sm{2}{39} \YfS{a}{e} \mu_{eb} &&+ \sm{2}{40} \YSYF{a}{e} \mu_{eb}  \nonumber\\
	+& \sm{2}{41} \tr{\m\yta\m\yte\yb\yte} &&+ \sm{2}{42}  \tr{\ya\ytb\m\yte\m\yte} \nonumber\\
	+& \sm{2}{43} \tr{\ya\mt\m\yte\yb\yte} &&+ \sm{2}{44} \tr{\ya\mt\yb\yte\m\yte}  \nonumber\\
	+& \sm{2}{45} \tr{\ya\ytb\ye\mt\m\yte} &&+ \sm{2}{46} \tr{\ya\mt\ye\mt\yb\yte}  \nonumber\\
	+& \sm{2}{47} \tr{\ya\mt\ye\ytb\m\yte} &&+ \sm{2}{48} \tr{\ya\ytb\m\mt\YF} \nonumber\\
	+& \sm{2}{49} \tr{\ya\mt\yb\mt\YF} &&+ \sm{2}{50} \tr{\ya\mt\m\ytb\YF}   \nonumber\\
	+& \sm{2}{51} \tr{\m\yta\yb\mt\YF} \,.
\end{align}
}

\begin{center} \textbf{Fermion mass coefficients} \end{center}

{\allowdisplaybreaks
At 1-loop:
\begin{align}
	\fc{1}{1} &= -6, & \fc{1}{2} &= 2, & \fc{1}{3} &= 1.
\end{align}

At 2-loop:
\begin{align}
	\fc{2}{1} &= 0, & \fc{2}{2} &= -3,  & \fc{2}{3} &= -\frac{97}{3}, & \fc{2}{4} &= \frac{11}{6}, & \fc{2}{5} &= \frac{5}{3}, & \fc{2}{6} &= 12, \nonumber \\
	\fc{2}{7} &= 0,  & \fc{2}{8} &= 6, & \fc{2}{9} &= 10, & \fc{2}{10} &= 6, & \fc{2}{11} &= 9,  & \fc{2}{12} &= -\frac{1}{2}, \nonumber \\
	\fc{2}{13} &= -\frac{7}{2}, & \fc{2}{14} &= -2, & \fc{2}{15} &= 2, & \fc{2}{16} &= 0, & \fc{2}{17} &= -2,  & \fc{2}{18} &= 0, \nonumber \\
	\fc{2}{19} &= -2, & \fc{2}{20} &= -\frac{1}{4}, & \fc{2}{21} &= -1, & \fc{2}{22} &= -\frac{3}{4} \,.
\end{align}
}

\begin{center} \textbf{Trilinear coefficients} \end{center}

{\allowdisplaybreaks
At 1-loop:
\begin{align}
	\tc{1}{1} &= -9, & \tc{1}{2} &= 3, & \tc{1}{3} &= \frac{3}{2}, & \tc{1}{4} &= -12\,.
\end{align}

At 2-loop:
\begin{align}
	\tc{2}{1} &= 6, & \tc{2}{2} &= 30,  & \tc{2}{3} &= 0, & \tc{2}{4} &= \frac{9}{2}, & \tc{2}{5} &= -\frac{143}{4}, & \tc{2}{6} &= \frac{11}{4}, \nonumber \\
	\tc{2}{7} &= \frac{10}{4}, & \tc{2}{8} &= -9,  & \tc{2}{9} &= 24, & \tc{2}{10} &= -\frac{9}{2}, & \tc{2}{11} &= -9, & \tc{2}{12} &= \frac{1}{4}, \nonumber \\
	\tc{2}{13} &= -3, & \tc{2}{14} &= -3,  & \tc{2}{15} &= 0, & \tc{2}{16} &= -36, & \tc{2}{17} &= -36, & \tc{2}{18} &= \frac{15}{2}, \nonumber \\
	\tc{2}{19} &= 0, & \tc{2}{20} &= -3,  & \tc{2}{21} &= 0, & \tc{2}{22} &= 0, & \tc{2}{23} &= 12, & \tc{2}{24} &= 6, \nonumber \\
	\tc{2}{25} &= -24, & \tc{2}{26} &= -24,  & \tc{2}{27} &= 6, & \tc{2}{28} &= 6, & \tc{2}{29} &= 0, & \tc{2}{30} &= 0, \nonumber \\
	\tc{2}{31} &= -\frac{3}{2}, & \tc{2}{32} &= -\frac{9}{4},  & \tc{2}{33} &= 24, & \tc{2}{34} &= 12, & \tc{2}{35} &= 12, & \tc{2}{36} &= 24, \nonumber \\
	\tc{2}{37} &= 12, & \tc{2}{38} &= 12\,.
\end{align}
}

\begin{center} \textbf{Scalar mass coefficients} \end{center}

{\allowdisplaybreaks
At 1-loop:
\begin{align}
	\sm{1}{1} &= -6, & \sm{1}{2} &= 1, & \sm{1}{3} &= 1, & \sm{1}{4} &= 1, & \sm{1}{5} &= -4\,, & \sm{1}{6} &= -2\,.
\end{align}

At 2-loop:
\begin{align}
	\sm{2}{1} &= 2, & \sm{2}{2} &= 10,  & \sm{2}{3} &= 0, & \sm{2}{4} &= 3, & \sm{2}{5} &= -\frac{143}{6}, & \sm{2}{6} &= \frac{11}{6}, \nonumber \\
	\sm{2}{7} &= \frac{10}{6}, & \sm{2}{8} &= -3,  & \sm{2}{9} &= 8, & \sm{2}{10} &= 8, & \sm{2}{11} &= -3, & \sm{2}{12} &= -3, \nonumber \\
	\sm{2}{13} &= \frac{1}{6}, & \sm{2}{14} &= -1,  & \sm{2}{15} &= -\frac{1}{2}, & \sm{2}{16} &= -2, & \sm{2}{17} &= 0, & \sm{2}{18} &= 0, \nonumber \\
	\sm{2}{19} &= -12, & \sm{2}{20} &= 5,  & \sm{2}{21} &= 0, & \sm{2}{22} &= -1, & \sm{2}{23} &= -1, & \sm{2}{24} &= 0, \nonumber \\
	\sm{2}{25} &= 0, & \sm{2}{26} &= 0,  & \sm{2}{27} &= 2, & \sm{2}{28} &= 4, & \sm{2}{29} &=-8 , & \sm{2}{30} &= -8, \nonumber \\
	\sm{2}{31} &= -4, & \sm{2}{32} &= -4,  & \sm{2}{33} &= 2, & \sm{2}{34} &= 4, & \sm{2}{35} &= 1, & \sm{2}{36} &= 0, \nonumber \\
	\sm{2}{37} &= 0, & \sm{2}{38} &= 0,  & \sm{2}{39} &= -1, & \sm{2}{40} &= -\frac{3}{2}, & \sm{2}{41} &= 4, & \sm{2}{42} &= 8, \nonumber \\
	\sm{2}{43} &= 8, & \sm{2}{44} &= 4,  & \sm{2}{45} &= 4, & \sm{2}{46} &= 4, & \sm{2}{47} &= 4, & \sm{2}{48} &= 4, \nonumber \\
	\sm{2}{49} &= 4, & \sm{2}{50} &= 2,  & \sm{2}{51} &= 2 \,.
\end{align}
}
\end{widetext}

\emph{Conclusions.}~--- We have presented in this Letter the general expressions of the two-loop \msbar~\bfuncs~for dimensionful parameters in the formalism of Poole \& Thomsen. This completes the set of dimensionless RGEs given in \cite{poole_constraints_2019} up to the respective loop orders 3-2-2 for gauge, Yukawa and quartic couplings. The full set of RGEs in this new formalism has been implemented in a new version of the PyR@TE software, PyR@TE 3 \cite{SARTORE2021107819, sartore_pyrate_git_2020}. We note finally that the procedure presented in this Letter would remain valid using different renormalization schemes, both in the non-supersymmetric and supersymmetric case.

\begin{acknowledgments}

We are grateful to Colin Poole and Anders Eller Thomsen for many useful discussions. We would also like to thank Ingo Schienbein for his comments and support in the preparation of this Letter.

This work was supported in part by the IN2P3 project ``Th\'eorie -- BSMGA''.

\vfill

\end{acknowledgments}

\bibliography{dimensionfulRGEs.bib}

\end{document}